# Study of Compact Stars with Buchdahl Potential in 5-D Einstein-Gauss-Bonnet Gravity

Manuel Malaver[1], Rajan Iyer[2], Israr Khan[3]

[1]Maritime University of the Caribbean, Department of Basic Sciences, Catia la Mar, Venezuela.
Email: **mmf.umc@gmail.com**

[2]Environmental Materials Theoretical Physicist, Department of Physical Mathematics Sciences Engineering Project Technologies, Engineeringinc International Operational Teknet Earth Global, Tempe, Arizona, United States of America
Email: engginc@msn.com

[3]Institute of Numerical Sciences, Kohat University of Science & Technology, Kohat, Pakistan .
Email: israrali@kust.edu.pk

**Abstract:** In this paper, we have presented a compact object model in the framework of Einstein-Gauss-Bonnet gravity (EGB) with a linear equation of state considering a metric potential proposed for Buchdahl (1959). The new obtained models satisfy all physical requirements of a physically reasonable stellar object. We analyzed the effect of the Gauss-Bonnet coupling constant $\alpha$ on the main physical characteristics of the model. We checked that the radial pressure, energy density and anisotropy are well defined and are regular in the interior of the star and are dependent of the values of the coupling constant.
**Keywords:** EGB gravity, Linear equation of state, Coupling constant, Compact object, Metric potential.

## 1. Introduction

Mathematical modeling within the framework of the general theory of relativity has been used to explain the behavior and structure of massive objects as neutron stars, quasars, black holes, pulsars and white dwarfs [1,2] and requires finding the exact solutions of the Einstein-Maxwell system [3]. A detailed and systematic analysis was carried out by Delgaty and Lake [4] which obtained several analytical solutions that can describe realistic stellar configurations.

It is very important to mention the pioneering works of Schwarzschild [5], Tolman [6], Oppenheimer and Volkoff [7] and Chandrasekhar [8] in the development of the first theoretical models of stellar objects. Schwarzschild [5] obtained interior solutions that allows describing a star with uniform density, Tolman [6] generated new solutions for static spheres of fluid, Oppenheimer and Volkoff [7] studied the gravitational equilibrium of neutron masses using the equation of state for a cold Fermi gas and general relativity and Chandrasekhar [8] produced new models of white dwarfs in presence of relativistic effects. Some of these results have been extended to higher dimensions and the dimensionality of space-time apparently influence the stability of these fluid spheres [9].

Recently, astronomical observations of compact objects have allowed new findings of neutron stars and strange stars that adjust to the exact solutions of the 4-D Einstein field equations and the data on mass maximum, redshift and luminosity are some of the most relevant characteristics for verifying the physical requirements of these models [10]. A great number of exact models from the Einstein-Maxwell field equations have been generated by Gupta and Maurya [11], Kiess [12], Mafa Takisa and Maharaj [13], Malaver and Kasmaei [14], Malaver [15,16], Ivanov [17] and Sunzu et al [18]. In the development of these models several forms of equations of state can be considered [19]. Komathiraj and Maharaj [20], Malaver [21], Bombaci [22], Thirukkanesh and Maharaj [23], Dey et al. [24] and Usov [25] assume linear equation of state for quark stars. Feroze and Siddiqui [26] considered a quadratic equation of state for the matter distribution and specified particular forms for the gravitational potential and electric field intensity. MafaTakisa and Maharaj [13] obtained new exact solutions to the Einstein-Maxwell system of equations with a polytropic equation of state. Thirukkanesh and Ragel [27] have obtained particular models of anisotropic fluids with polytropic equation of state which are consistent with the reported experimental observations. Malaver [28] generated new exact solutions to the Einstein-Maxwell system considering Van der Waals modified equation of state with polytropic exponent. Tello-Ortiz et al. [29] found an anisotropic fluid sphere solution of the Einstein-Maxwell field equations with a modified version of the Chaplygin equation of state.

The analysis of compact objects with anisotropic matter distribution is very important, because that the anisotropy plays a significant role in the studies of relativistic spheres of fluid [30-42]. Anisotropy is defined as $\Delta = p_t - p_r$ where $p_r$ is the radial pressure and $p_t$ is the tangential pressure. The existence of solid core, presence of type 3A superfluid [43], magnetic field, phase transitions, a pion condensation and electric field [25] are most important reasonable facts that explain the presence of tangential pressures within a star. Many astrophysical objects as X-ray pulsar, Her X-1, $4U$1820-30 and SAXJ1804.4-3658 have anisotropic pressures. Bowers and Liang [42] include in the equation of hydrostatic equilibrium the case of local anisotropy. Bhar et al. [44] have studied the behavior of relativistic objects with locally anisotropic matter distribution considering the Tolman VII form for the gravitational potential with a linear relation between the energy density and the radial pressure. Malaver [45-46], Feroze and Siddiqui [26,47] and Sunzu et al.[18] obtained solutions of the Einstein-Maxwell field equations for charged spherically symmetric space-time by assuming anisotropic pressure.

The behavior and dynamics of the gravitational field can be extended to higher dimensions [48]. The history of higher dimensions goes back to the work done by Kaluza [49] and Klein [50] who introduced the concept of extra dimensions in addition to the usual four dimensions (4-D) to unify gravitational and electromagnetic interactions. In general theory of relativity, the results obtained in four dimensions can be generalized in higher dimensional context and study the effects due to incorporation of extra space-time dimensions [51]. Within this framework, a very useful and fruitful generalization is the Einstein-Gauss-Bonnet gravity, which has generated a lot of interest among researchers and has been influenced by many scientists working in this field [52]. The modeling of compact objects in EGB gravity has shown that some physical variables are modified when they are compared to their 4-D counterparts, but the condition of the Schwarzschild constant density sphere has been demonstrated in EGB gravity [10]. Recently, Bhar et al. [53] performed a comparative study of compact objects in five dimensions (5-D) between EGB gravity and classical general relativity theory and found that many features as stability, causality and energy conditions remain unaffected in the stellar interior. Per publications with quantum astrophysical sciences [54-57] understanding of a unified knowledge of physics characterizing universe mechanisms with quantum astrophysics have provided key to quantum as well as astrophysical studies of interior of stellar galaxies.

In this work, we have used the metric potential proposed for Buchdahl [58] to generate some stellar models with anisotropic matter distribution in EGB gravity. The system of field equations has been solved to obtain analytic solutions which are physically acceptable. The paper is organized as follows: In Section.2, we present the framework of EGB gravity. The modified Einstein-Maxwell field equations with the Gauss-Bonnet coupling constant are presented in Section.3. With the Buchdahl ansatz, we generate some models of an anisotropic star with a linear equation of state within EGB gravity in Section.4. In Section. 5, physical requirements for the new models are described. In Section.6, a physical analysis of the new solutions is performed. In final Section, we conclude.

## 2. Einstein-Gauss-Bonnet Gravity

The Gauss-Bonnet action in n (n≥5)-dimensional spacetime can be written as

$$S = \int \sqrt{-g}\, d^n x \left[ \frac{1}{2k_n^2}\left(R + \alpha L_{GB}\right) \right] + S_{matter} \tag{1}$$

where $\alpha$ is the Gauss-Bonnet coupling constant. The strength of the action $L_{GB}$ lies in the fact that despite the Lagrangian being quadratic in the Ricci tensor, Ricci scalar and the Riemann tensor, the equations of motion turn out to be second order quasi-linear which are compatible with Einstein's theory of gravity [53,54].

The EGB field equations may be written as

$$G_{ab} + \alpha H_{ab} = T_{ab} \tag{2}$$

where $G_{ab}$ represents the Einstein tensor, $T_{ab}$ is the total energy-momentum tensor and the Lanczos tensor $H_{ab}$ is given by

$$H_{ab} = 2(RR_{ab} - 2R_{ac}R_b^c - 2R^{cd}R_{acbd} + R_a^{cde}R_{bcde}) - \frac{1}{2}g_{ab}L_{GB} \tag{3}$$

where the Lovelock term has the form

$$L_{GB} = R^2 + R_{abcd}R^{abcd} - 4R_{cd}R^{cd} \tag{4}$$

### 3. Field Equations

The 5-dimensional line element for a static spherically symmetric space-time takes the form

$$ds^2 = -e^{2\nu(r)}dt^2 + e^{2\lambda(r)}dr^2 + r^2(d\theta^2 + \sin^2\theta d\phi^2 + \sin^2\theta\sin^2\phi d\psi^2) \tag{5}$$

where the metric functions $e^\nu$ and $e^\lambda$ are the gravitational potentials. By considering the commoving fluid velocity as $u^a = e^{-\nu}\delta_0^a$, the EGB field equations (2) reduce to

$$\rho = \frac{3}{e^{4\lambda}r^3}\left(4\alpha\lambda' + re^{2\lambda} - re^{4\lambda} - r^2 e^{2\lambda}\lambda' - 4\alpha e^{2\lambda}\lambda'\right) \tag{6}$$

$$p_r = \frac{3}{e^{4\lambda}r^3}\left(-re^{4\lambda} + \left(r^2\nu' + r + 4\alpha\nu'\right)e^{2\lambda} - 4\alpha\nu'\right) \tag{7}$$

$$p_t = \frac{1}{e^{4\lambda}r^2}\left(-e^{4\lambda} - 4\alpha\nu'' + 12\alpha\nu'\lambda' - 4\alpha(\nu')^2\right) +$$
$$\frac{1}{e^{2\lambda}r^2}\left(1 - r^2\nu'\lambda' + 2r\nu' - 2r\lambda' + r^2(\nu')^2\right) +$$
$$\frac{1}{e^{2\lambda}r^2}\left(r^2\nu'' - 4\alpha\nu'\lambda' + 4\alpha(\nu')^2 + 4\alpha\nu''\right) \tag{8}$$

Here primes means a derivation with respect to the radial coordinates $r$ and $\rho$ is the energy density, $p_r$ is the radial pressure and $p_t$ is the tangential pressure. With the transformations $x = cr^2$, $Z(x) = e^{-2\lambda}$ and $y^2(x) = e^{2\nu}$ suggested by Durgapal and Bannerji [59] and with $c > 0$ as arbitrary constant, the field equations (6)-(8) can be written as follows

$$\frac{\rho}{c} = -3\dot{Z} - \frac{3(Z-1)(1-4\beta\dot{Z})}{x} \tag{9}$$

$$\frac{p_r}{c} = \frac{3(Z-1)}{x} + \frac{6Z\dot{y}}{y} - \frac{6\beta(Z-1)Z\dot{y}}{xy} \qquad (10)$$

$$\frac{p_t}{c} = 4Z\left[\beta(1-Z)+x\right]\frac{\ddot{y}}{y} + \left[\frac{2\beta Z(1-Z)}{x} - 2(x+\beta)\dot{Z} + 6Z - 2\beta Z\dot{Z}\right]\frac{\dot{y}}{y} + 2\left[\frac{Z-1}{x} + \dot{Z}\right] \qquad (11)$$

where $\beta = 4\alpha c$ contains the Gauss-Bonnet coupling constant $\alpha$ and dots denote differentiation with respect to $x$.

In this paper, we assume the following equation of state

$$p_r = m\rho \qquad (12)$$

where $m$ is an arbitrary constant.

## 4. The New Models

In this research, we take the form of the metric potential $Z(x)$ proposed for Buchdahl [58].

$$Z(x) = \frac{K+x}{K(1+x)} \qquad (13)$$

where $K$ is a parameter related to the geometry of the star. This potential is regular at the stellar center and well behaved in the interior of the sphere. Using $Z(x)$ in equation (9), we obtain

$$\rho = c\frac{\left[3K(K-1)(x+1) + 3K(K-1)(x+1)^2 + 12\beta(K-1)^2\right]}{K^2(x+1)^3} \qquad (14)$$

Substituting the equation (14) in the expression of the linear equation of state for the radial pressure (12), we have

$$p_r = mc\frac{\left[3K(K-1)(x+1) + 3K(K-1)(x+1)^2 + 12\beta(K-1)^2\right]}{K^2(x+1)^3} \qquad (15)$$

With $Z(x)$ and (15) in equation (10), we can written

$$\frac{\dot{y}}{y} = \frac{3K(K-1)(1+x)}{6K(K+x)(1+x) + 6\beta(K-1)(K+x)} + m\frac{\left[3K(K-1)(x+1) + 3K(K-1)(x+1)^2 + 12\beta(K-1)^2\right]}{(1+x)\left[6K(K+x)(1+x) + 6\beta(K-1)(K+x)\right]}$$

(16)

Integrating equation (16) we obtain

$$y(x) = c_1 (Kx + K\beta + K - \beta)^A (K + x)^B (1 + x)^C \qquad (17)$$

where $c_1$ is the constant of integration. The constants $A$, $B$ and $C$ are given by

$$A = -\frac{(K\beta m + K\beta + 3Km - \beta m - \beta)}{2(K - \beta)} \qquad (18)$$

$$B = \frac{K^2 m + K^2 - 2Km + 4\beta m - K}{2(K - \beta)} \qquad (19)$$

$$C = 2m \qquad (20)$$

For the metric functions $e^{2\lambda}$ and $e^{2\nu}$, we have

$$e^{2\lambda} = \frac{K(1+x)}{K+x} \qquad (21)$$

$$e^{2\nu} = c_1^2 (Kx + K\beta + K - \beta)^{2A} (K + x)^{2B} (1 + x)^{2C} \qquad (22)$$

and the anisotropy can be written as

$$\Delta = p_t - p_r = \frac{4xc(K+x)(\beta x - \beta + Kx + K)}{K^2(1+x)^2} \left[ \frac{(A^2 - A^2)K^2}{(\beta K + Kx - \beta + K)^2} + \frac{2AKB}{(\beta K + Kx - \beta + K)(K+x)} + \frac{2AKC}{(\beta K + Kx - \beta + K)(1+x)} + \frac{B^2 - B}{(K+x)^2} + \frac{2BC}{(K+x)(1+x)} + \frac{C^2 - C}{(1+x)^2} \right]$$

$$+ \left[ \frac{4\beta(1-K)(K+x)}{K^2(1+x)^2} - \frac{2(1-K)(x+\beta)}{K(1+x)^2} - \frac{2\beta(1-K)(K+x)}{K^2(1+x)^3} \right] \left[ \frac{AK}{\beta K + Kx - \beta + K} + \frac{B}{K+x} + \frac{C}{1+x} \right]$$

$$+ \frac{2c(1-K)}{K(1+x)^2} - \frac{2c(1-K)}{K(1+x)}$$

(23)

## 5. Physical Acceptability in EGB Gravity

For a model to be physically acceptable in EGB gravity, the following conditions should be satisfied [27,53]:

(i) The metric potentials $e^{2\lambda}$ and $e^{2\nu}$ assume finite values throughout the stellar interior and are singularity-free at the center $r=0$.

(ii) The energy density $\rho$ and the radial pressure $p_r$ should be positive inside the star.

(iii) The anisotropy is zero at the center $r=0$, i.e. $\Delta(r=0) = 0$.

(iv) The energy density and radial pressure are decreasing functions with the radial parameter, i.e. $\frac{dp_r}{dr} \leq 0$ and $\frac{d\rho}{dr} \leq 0$ both in EGB gravity.

(v) Any physically acceptable model must satisfy the causality condition, that is, for the radial sound speed $v_{sr}^2 = \frac{dp_r}{d\rho}$, we should have $0 \leq v_{sr}^2 \leq 1$.

(vi) The boundary of the star defined by $r=R$ should be matched with the Einstein –Gauss-Bonnet- Schwarzschild exterior solution given by

$$ds^2 = -F(r)dt^2 + \frac{dr^2}{F(r)} + r^2(d\theta^2 + \sin^2\theta d\phi^2 + \sin^2\theta \sin^2\phi d\psi^2) \tag{35}$$

where $R$ is the radius of the star and

$$F(r) = 1 + \frac{r^2}{4\alpha}\left(1 - \sqrt{1 + \frac{8M\alpha}{r^4}}\right) \tag{36}$$

In Equation (36) $M$ is associated with the gravitational mass of the hypersphere

## 6. Physical Features of the New Models

In order to obtain the parameters $A$, $B$, $C$, $K$ that describe the model and ensure the matching conditions is used the first fundamental form that consist in the continuity of the metric functions and their derivatives across the boundary $r=R$ as follows

$$\frac{K(1+cR^2)}{K+cR^2} = 1 + \frac{R^2}{4\alpha}\left(1 - \sqrt{1 + \frac{8\alpha M}{R^4}}\right) \tag{37}$$

$$c_1(KcR^2 + K\beta + K - \beta)^A (K + cR^2)^B (1 + cR^2)^C = 1 + \frac{R^2}{4\alpha}\left(1 - \sqrt{1 + \frac{8\alpha M}{R^4}}\right) \tag{38}$$

$$2c_1(KcR^2 + K\beta + K - \beta)^A (K + cR^2)^B (1 + cR^2)^C \left[\frac{AKc}{KcR^2 + K\beta + K - \beta} + \frac{Bc}{K + cR^2} + \frac{Cc}{1 + cR^2}\right]$$

$$= -\frac{1}{2\alpha} \frac{\left(1 - \sqrt{1 + \frac{8\alpha M}{R^4}}\right)}{\sqrt{1 + \frac{8\alpha M}{R^4}}}$$

$$\tag{39}$$

and from the second fundamental form, we obtain

$$R^4 + \frac{3}{c}R^2 + \frac{2}{c^2} + \frac{4\beta(K-1)}{Kc^2} = 0 \tag{40}$$

The metric potentials $e^{2\lambda}$ and $e^{2\nu}$ have finite values and remain positive throughout the stellar interior. At the center $e^{2\lambda(0)} = 1$, $e^{2\nu(0)} = c_1^2(K\beta + K - \beta)^{2A} K^{2B}$. We show that in $r=0$, $\left(e^{2\lambda(r)}\right)'_{r=0} = \left(e^{2\nu(r)}\right)'_{r=0} = 0$ and it is verified that the gravitational potentials are regular at the center.

The energy density and radial pressure are positive and well behaved in the stellar interior. The obtained central density are $\rho(r=0) = \frac{c[6K(K-1) + 12\beta(K-1)^2]}{K^2}$ and $p_r(r=0) = mc\frac{[6K(K-1) + 12\beta(K-1)^2]}{K^2}$, both positive if $m, K, \beta > 0$.

In the surface of the fluid sphere $r=R$, we have $p_r(r=R) = 0$ and is obtained for the radius of the star

$$R = \frac{\sqrt{-2Kc\left(3K - \sqrt{-16K^2\beta + K^2 + 16K\beta}\right)}}{2cK} \tag{41}$$

Differentiating Eq. (14) and Eq. (15), the expressions for density and radial pressure gradient are given by

$$\frac{d\rho}{dr} = \frac{c^2 r\left[6K(K-1)+12K(K-1)(1+cr^2)\right]}{K^2(1+cr^2)^3} - \frac{6c^2 r\left[3K(K-1)(1+cr^2)+3K(K-1)(1+cr^2)^2+12\beta(K-1)^2\right]}{K^2(1+cr^2)^4}$$

(42)

$$\frac{dp_r}{dr} = m\frac{c^2 r\left[6K(K-1)+12K(K-1)(1+cr^2)\right]}{K^2(1+cr^2)^3} - m\frac{6c^2 r\left[3K(K-1)(1+cr^2)+3K(K-1)(1+cr^2)^2+12\beta(K-1)^2\right]}{K^2(1+cr^2)^4}$$

(43)

For the physically acceptability of the model of star, it should satisfy the causality condition, i.e. $0 \leq v_{sr}^2 \leq 1$ and the following energy conditions must be satisfied: *NEC* ( Null energy conditions, *WEK* ( Weak energy conditions), *SEC* (Strong energy conditions) [54,55] expressed through the inequalities

$$NEC: \rho \geq 0 \qquad (44)$$

$$WEC: \rho + p_r \geq 0 \qquad (45)$$

$$SEC: \rho + p_r + 2p_t \geq 0 \qquad (46)$$

In the Table 1 shows the values of the parameters *K*, *m* and the stellar radius *R* when the coupling constant *α* not vary.

**Table 1**. Parameters *K*, *m* and *R* with *α* constant

| α | K | m | R (Km) |
|---|---|---|---|
| 20 | 0.2 | 1/3 | 5.85 |
| 20 | 0.3 | 1/3 | 5.08 |
| 20 | 0.4 | 1/3 | 4.52 |
| 20 | 0.5 | 1/3 | 4.05 |

The figures 1, 2, 3, 4, 5, 6 and 7 present the dependence of $\rho$, $p_r$, $\frac{d\rho}{dr}$, $\frac{dp_r}{dr}$, $\Delta$, *WEC* and *SEC* with the radial coordinate for the parameters given in the Table 1.

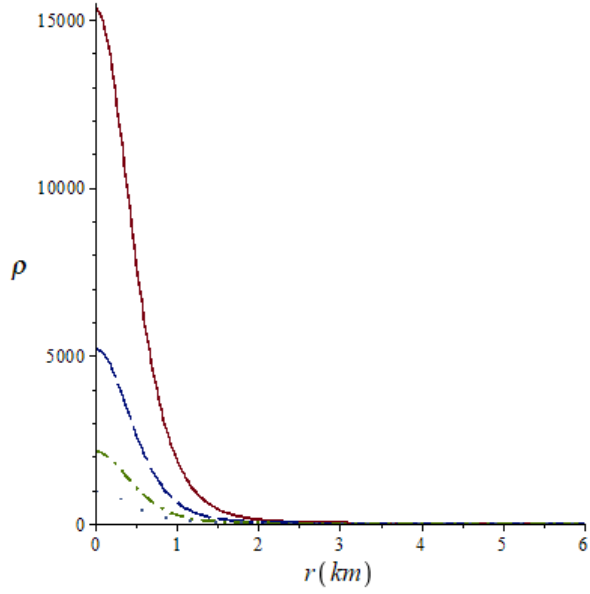

*Figure 1. Energy density against radial coordinate for the parameters given in Table 1. It has been considered that K=0.2 (solid line); K=0.3 (long-dash line); K=0.4 (dashdot line); K=0.5 ( spacedot line).*

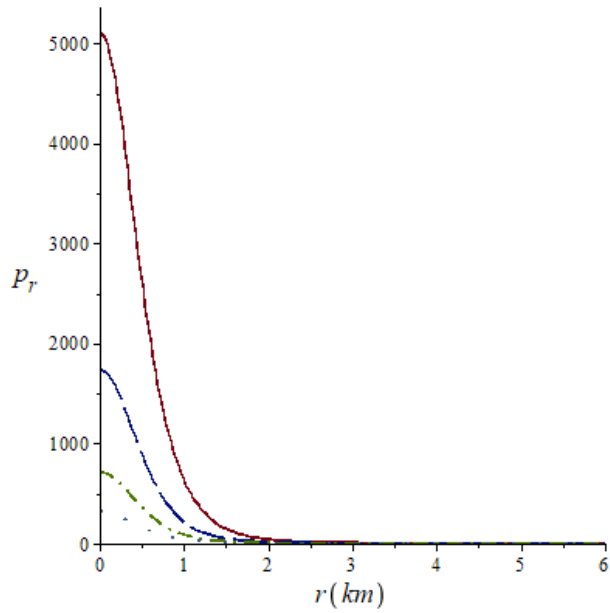

*Figure 2. Radial pressure against radial coordinate for the parameters given in Table 1. It has been considered that K=0.2 (solid line); K=0.3 (long-dash line); K=0.4 (dashdot line); K=0.5 ( spacedot line).*

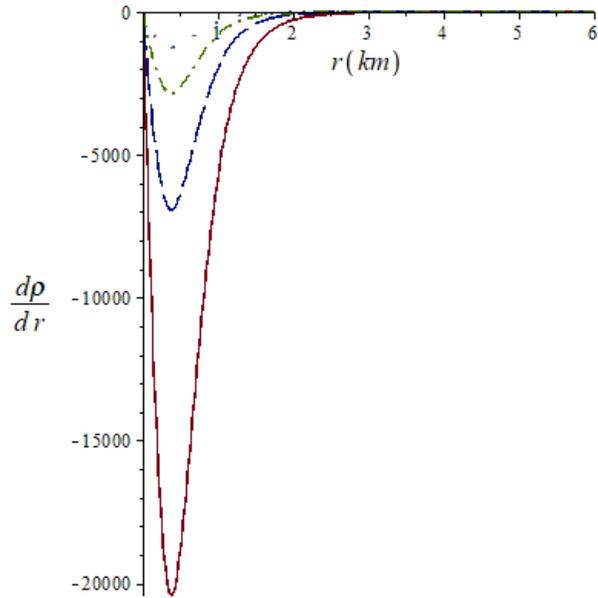

*Figure 3.* Density gradient against radial coordinate for the parameters given in Table 1. It has been considered that K=0.2 (solid line); K=0.3 (long-dash line); K=0.4 (dashdot line); K=0.5 (spacedot line).

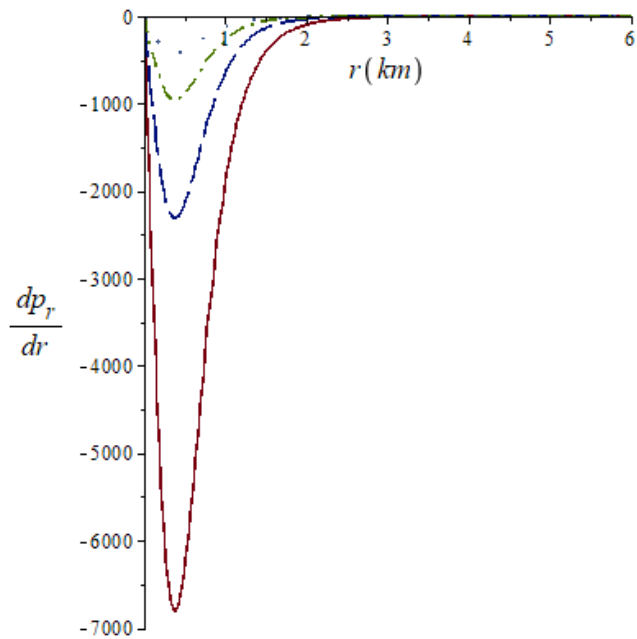

*Figure 4.* Radial pressure gradient against radial coordinate for the parameters given in Table 1. It has been considered that K=0.2 (solid line); K=0.3 (long-dash line); K=0.4 (dashdot line); K=0.5 (spacedot line).

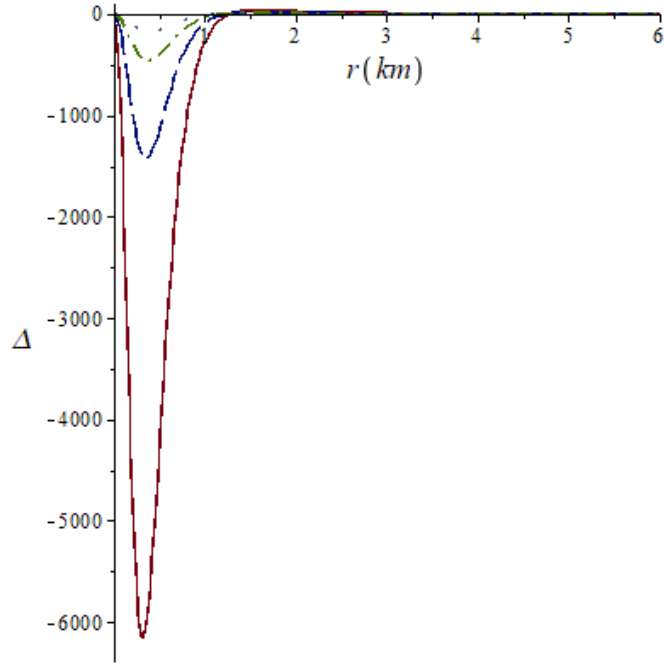

*Figure 5.* Anisotropy against radial coordinate for the parameters given in Table 1. It has been considered that K=0.2 (solid line) ; K=0.3 (long-dash line); K=0.4 (dashdot line); K=0.5 (spacedot line).

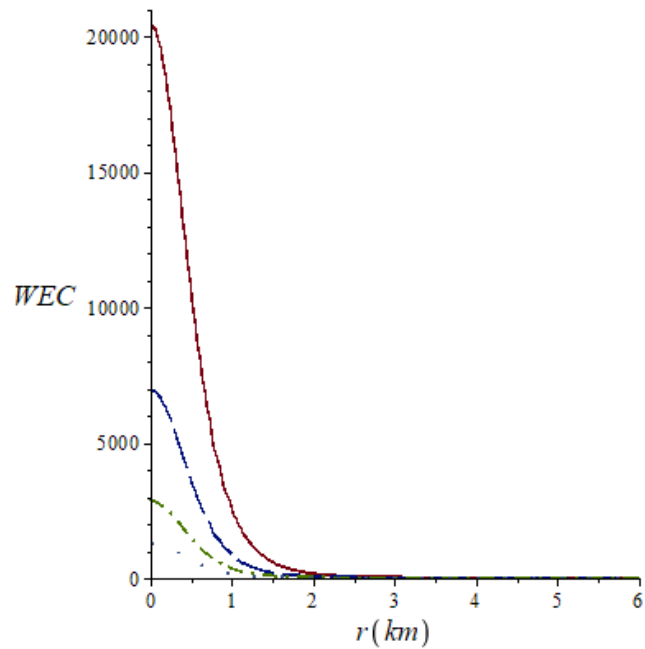

*Figure 6.* WEC against radial coordinate for the parameters given in Table 1. It has been considered that K=0.2 (solid line) ; K=0.3 (long-dash line); K=0.4 (dashdot line); K=0.5 (spacedot line).

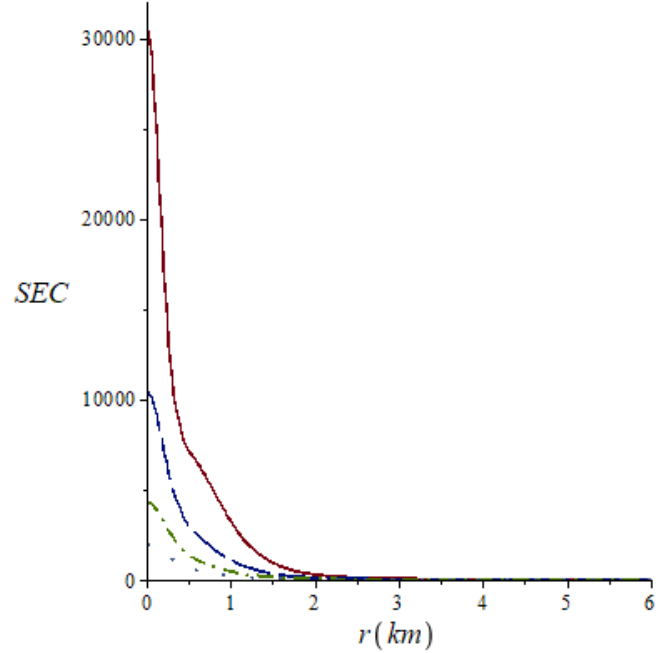

*Figure 7. SEC against radial coordinate for the parameters given in Table 1. It has been considered that K=0.2 (solid line) ;K=0.3 (long-dash line); K=0.4 (dashdot line); K=0.5 (spacedot line).*

In the Figure 1 is shown that the energy density remains positive, continuous and is monotonically decreasing function throughout the stellar interior for all values of *K*. The radial pressure showed the same behavior by the energy density, that is, it is growing within the star and vanishes at a greater radial distance, but takes the lower values when *K* is increased as shown in Figure 2. In Figure 3 it is noted that for the radial variation of energy density gradient $\frac{d\rho}{dr} < 0$ in the four cases studied. Again, according to Figure 4, the profile of $\frac{dp_r}{dr}$ shows that radial pressure gradient is negative inside the star. The anisotropy is plotted in Figure 5 and it shows that vanishes at the centre of the star, i.e. $\Delta(r=0) =0$ [41,48]. We can also note that $\Delta$ admits higher values when *K* increases. The profiles of energy conditions *WEC* and *SEC* are graphically shown in Figures 6 and 7. These profiles indicate that the solution presented in this work is physically viable. In the table 2 shows of the values of the parameters *K*, *m* and the stellar radius *R* when the constant *K* not vary.

**Table 2**. Parameters *m*, *α* and *R* with *K* constant

| α | K | m | R (Km) |
|---|---|---|---|
| 10 | 0.5 | 1/3 | 3.34 |
| 20 | 0.5 | 1/3 | 4.05 |
| 30 | 0.5 | 1/3 | 4.52 |
| 40 | 0.5 | 1/3 | 4.88 |

The figures 8, 9, 10, 11, 12, 13 and 14 present the dependence of $\rho$, $p_r$, $\frac{d\rho}{dr}$, $\frac{dp_r}{dr}$, $\Delta$, WEC and SEC with the radial coordinate for the parameters given in the Table 2.

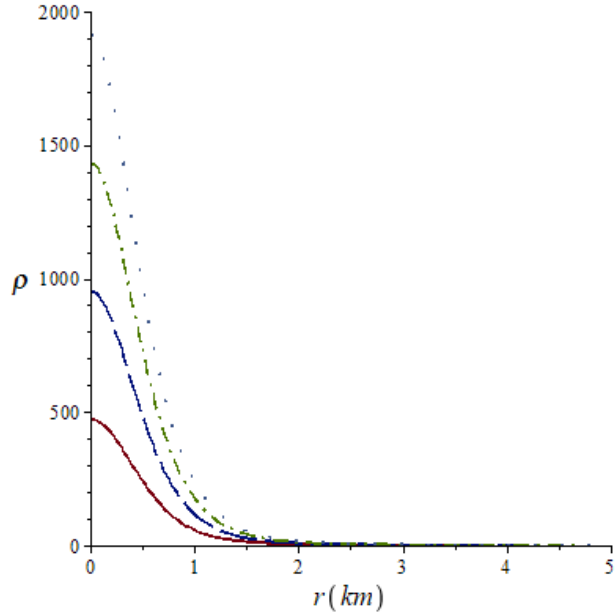

*Figure 8.* Energy density against radial coordinate for the parameters given in Table 2. It has been considered that $\alpha=10$ (solid line); $\alpha=20$ (long-dash line); $\alpha=30$ (dashdot line); $\alpha=40$ (spacedot line).

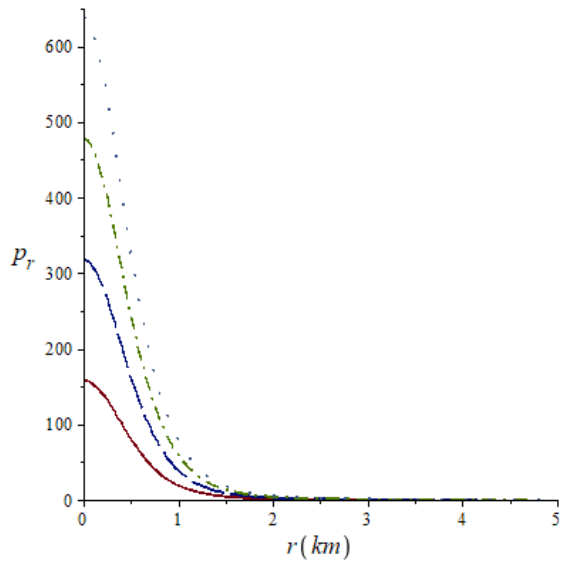

*Figure 9.* Radial pressure against radial coordinate for the parameters given in Table 2. It has been considered that $\alpha=10$ (solid line); $\alpha=20$ (long-dash line); $\alpha=30$ (dashdot line); $\alpha=40$ (spacedot line).

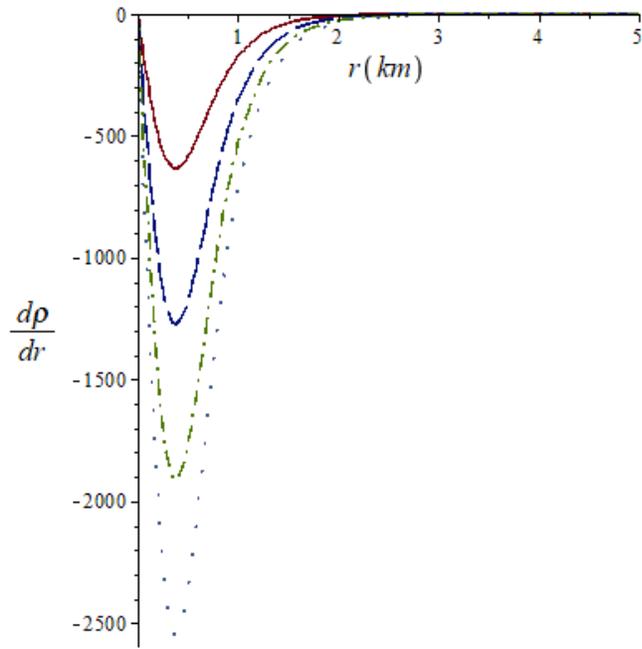

*Figure 10.* Density gradient against radial coordinate for the parameters given in Table 2. It has been considered that α=10 (solid line); α=20 (long-dash line); α=30 (dashdot line); α=40 ( spacedot line).

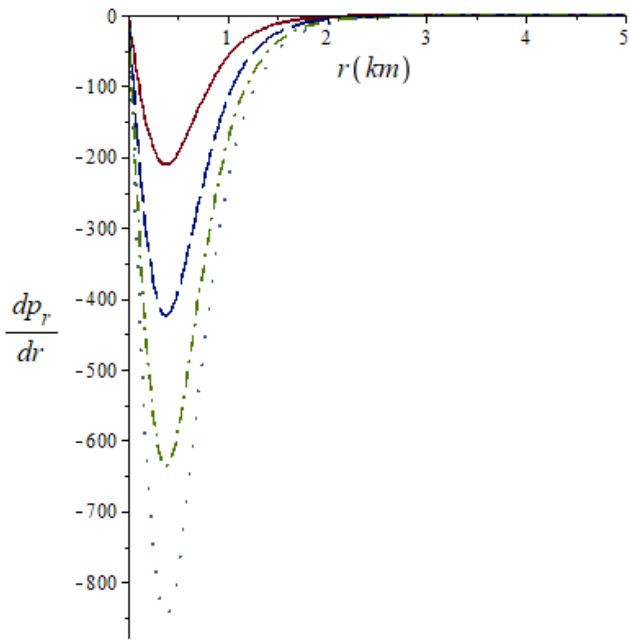

*Figure 11.* Radial pressure gradient against radial coordinate for the parameters given in Table 2. It has been considered that α=10 (solid line); α=20 (long-dash line); α=30 (dashdot line); α=40 ( spacedot line).

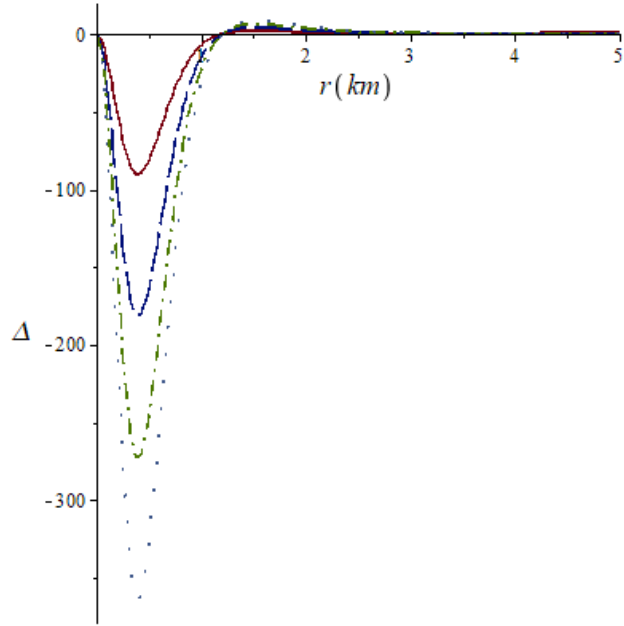

*Figure 12.* Anisotropy against radial coordinate for the parameters given in Table 2. It has been considered that α=10 (solid line); α=20 (long-dash line); α=30 (dashdot line); α=40 ( spacedot line).

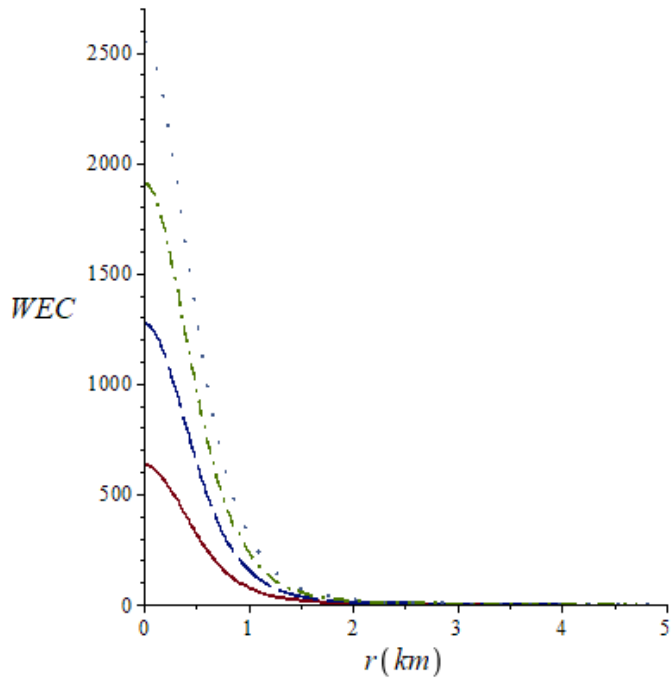

*Figure 13.* WEC against radial coordinate for the parameters given in Table 2. It has been considered that α=10 (solid line); α=20 (long-dash line); α=30 (dashdot line); α=40 ( spacedot line).

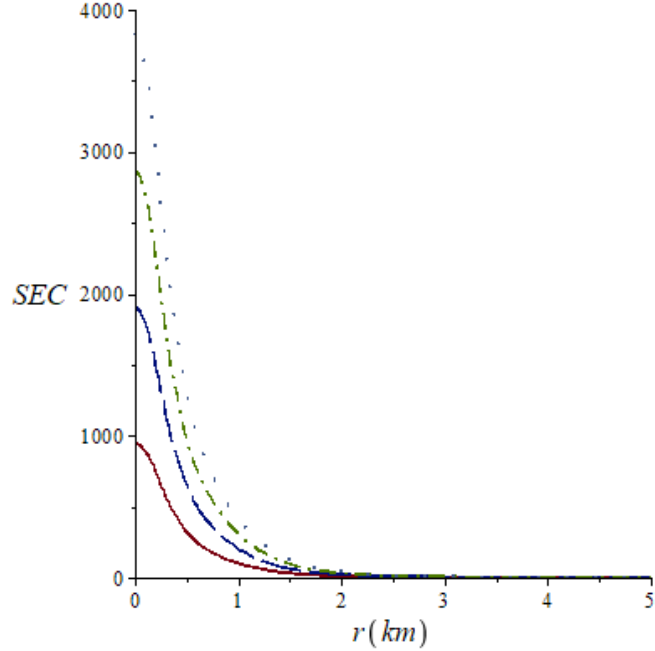

*Figure 14.* SEC against radial coordinate for the parameters given in Table 2. It has been considered that α=10 (solid line); α=20 (long-dash line); α=30 (dashdot line); α=40 ( spacedot line).

As in the case with α= constant, in the figure 8 the energy density is continuous and is monotonically decreasing inside the star and increases for higher values of α. With K=constant, the radial pressure always is positive throughout the stellar interior and vanishes at a finite radial distance and its results are shown in Figure 9. Again, the radial pressure increases when α takes higher values. The radial variation of energy density gradient has been shown in Figure 10, in which it is note that $\frac{d\rho}{dr} < 0$. In the Figure 11, it is also verified that the gradient $\frac{dp_r}{dr}$ is negative inside the star. In the Figure 12, the anisotropy $\Delta$ is zero at the center *r=0* and its value decreases when α takes higher values. Figures 13 and 14 again show that the energy conditions given by the inequalities (45) and (46) guarantee the physical viability of this model.

## 7. Conclusion

In this paper, we have generated new models of compact stars within the framework of Einstein-Gauss-Bonnet gravity so it is feasible to obtain models that describe real compact objects such as white dwarfs and neutron stars. With the use of Buchdahl ansatz for the gravitational potentials and with a linear equation of state, we are able to produce a new class of exact solution of the EGB field equations. We show that the developed configuration obeys the rigorous conditions required for the physical viability of the stellar model. A graphical analysis shows that the radial pressure, energy density and anisotropy are regular at the origin and well behaved in the interior. It is to be noted in EGB gravity that the coupling constant $α$ has non-negligible effects on the physical quantities such as energy density and radial pressure of the star which increases with an increase in $α$ when $K$ remain constant. The new solutions match smoothly with the exterior of the Einstein –Gauss-Bonnet- Schwarzschild at the boundary $r=R$, because matter variables and the gravitational potentials of this work are consistent with the physical analysis of these stars. As expected, the matching conditions require that the radial pressure vanishes at some finite radius of the stellar object and this defines the boundary of the star.

It is important to note that the analysis of the behavior of energy conditions with respect to the radial coordinate in the stellar interior shows that the obtained model is well satisfied in the context of EGB gravity for various values of $α$.

With earlier publications, these are consistent with quantum astrophysical sciences that will help to link micro macro aspects of intricate details of the workings of the cosmos compact stellar interior with quantum level atomistics that are prevalent in immediate vicinity of our environmental surroundings all the way to the astro-universe.

## References

[1] Kuhfitting, P.K.(2011). Some remarks on exact wormhole solutions, Adv. *Stud. Theor. Phys*. 2011, 5, 365.

[2] Bicak, J. Einstein equations: exact solutions, *Encyclopaedia of Mathematical Physics*. 2006,2, 165.

[3] Malaver, M. Black Holes, Wormholes and Dark Energy Stars in General Relativity. Lambert Academic Publishing, Berlin. ISBN: 978-3-659-34784-9, 2013.

[4] Delgaty.;M.S.R.; Lake, K. Physical Acceptability of Isolated, Static, Spherically Symmetric, Perfect Fluid Solutions of Einstein's Equations. *Comput. Phys. Commun.* 1998,115, 395.

[5] Schwarzschild, K. Uber das Gravitationsfeld einer Kugel aus inkompressibler Flussigkeit nach der Einsteinschen Theorie. *Math.Phys.Tech*, 1916, 424-434.

[6] Tolman, R.C. Static Solutions of Einstein's Field Equations for Spheres of Fluid. *Phys. Rev*. 1939, 55, 364-373.

[7] Oppenheimer, J.R. and Volkoff, G. On Massive Neutron Cores. *Phys. Rev*. 1939, 55, 374-381.

[8] Chandrasekahr, S. The Maximum mass of ideal white dwarfs. *Astrophys. J.* 1931, 74, 81-82.

[9] Bhar P.; Rahaman F.; Ray S.; Chatterjee V. Possibility of higher-dimensional anisotropic compact star. *Eur. Phys. J. C.* 2015; 75(5):190.

[10] Bhar, P.; Govender, M. Charged compact star model in Einstein-Maxwell-Gauss-gravity. *Astrophys Space Sci*. 2019, 364,186. https://doi.org/10.1007/s10509-019-3675-0

[11] Gupta, Y.K.; Maurya, S.K. A class of charged analogues of Durgapal and Fuloria superdense star. *Astrophys. Space Sci*. 2011, 331, 135-144.

[12] Kiess, T.E. Exact physical Maxwell-Einstein Tolman-VII solution and its use in stellar models . *Astrophys. Space Sci.* 2012, 339, 329-338.

[13] Mafa Takisa, P.; Maharaj, S.D. Some charged polytropic models. *Gen.Rel.Grav.* 2013, 45, 1951-1969.

[14] Malaver, M.; Kasmaei, H.D. Relativistic stellar models with quadratic equation of state. *International Journal of Mathematical Modelling & Computations*. 2020, 10, 111-124.

[15] Malaver, M. New Mathematical Models of Compact Stars with Charge Distributions. *International Journal of Systems Science and Applied Mathematics.* 2017, 2, 93-98.

[16] Malaver, M. Generalized Nonsingular Model for Compact Stars Electrically Charged. *World Scientific News.* 2018, 92, 327-339.

[17] Ivanov, B.V. Static charged perfect fluid spheres in general relativity. *Phys. Rev.D65*. 2002, 104011.

[18] Sunzu, J.M.; Maharaj, S.D.; Ray, S. Quark star model with charged anisotropic matter. *Astrophysics. Space.Sci.* 2014, 354, 517-524.

[19] Sunzu,J.M. Realistic Polytropic Models for Neutral Stars with Vanishing Pressure Anisotropy. *Global Journal of Science Frontier Research: A Physics and Space Science.* 2018, 18, ISSN 2249-4626. https://journalofscience.org/index.php/GJSFR/article/view/2168

[20] Komathiraj, K.; Maharaj, S.D. Analytical models for quark stars. *Int. J. Mod. Phys.* 2007, D16, 1803-1811.

[21] Malaver, M. Analytical models for compact stars with a linear equation of state. *World Scientific News*, 2016, 50, 64-73.

[22] Bombaci, I. Observational evidence for strange matter in compact objects from the x- ray burster 4U 1820-30, *Phys. Rev.* 1997, C55, 1587- 1590.

[23] Thirukkanesh, S. and Maharaj, S.D. Charged anisotropic matter with a linear equation of state. *Class. Quantum Gravity.* 2008, 25, 235001.

[24] Dey, M.; Bombaci, I.;, Dey, J.; Ray, S.; Samanta, B.C. Strange stars with realistic quark vector interaction and phenomenological density-dependent scalar potential, *Phys. Lett.* 1998, B438, 123-128.

[25] Usov, V. V. Electric fields at the quark surface of strange stars in the color-flavor locked phase. *Phys. Rev. D.* 2004, 70, 067301.

[26] Feroze, T.; Siddiqui, A. Charged anisotropic matter with quadratic equation of state. *Gen. Rel. Grav*. 2011, 43, 1025-1035.

[27] Thirukkanesh, S.; Ragel, F.C. Exact anisotropic sphere with polytropic equation of state. *PRAMANA-Journal of physics.* 2012, 78, 687-696.

[28] Malaver, M. Analytical model for charged polytropic stars with Van der Waals Modified Equation of State. *American Journal of Astronomy and Astrophysics.* 2013, 1, 37-42.

[29] Tello-Ortiz, F.; Malaver, M.; Rincón, A.; Gomez-Leyton, Y. Relativistic Anisotropic Fluid Spheres Satisfying a Non-Linear Equation of State. *Eur. Phys. J. C.* 2020, *80*, 371.

[30] Esculpi, M.; Malaver, M.; Aloma, E. A Comparative Analysis of the Adiabatic Stability of Anisotropic Spherically Symmetric solutions in General Relativity. *Gen. Relat.Grav.* 2007, 39, 633-652.

[31] Cosenza M.; Herrera L.; Esculpi M.; Witten L. Evolution of radiating anisotropic spheres in general relativity. *Phys.Rev.*1982, *D* 25, 2527-2535.

[32] Herrera L. Cracking of self-gravitating compact objects. *Phys. Lett.* 1992, *A*165, 206-210.

[33] Herrera L.; Nuñez L. Modeling 'hydrodynamic phase transitions' in a radiating spherically symmetric distribution of matter. *The Astrophysical Journal.* 1989, 339, 339-353.

[34] Herrera L.; Ruggeri G. J.; Witten L. Adiabatic Contraction of Anisotropic Spheres in General Relativity. *The Astrophysical Journal.* 1979, 234, 1094-1099.

[35] Herrera L.; Jimenez L.; Leal L.; Ponce de Leon J.; Esculpi M.; Galina V. Anisotropic fluids and conformal motions in general relativity. *J. Math. Phys. 1984,* 25, 3274.

[36] Malaver, M. Quark Star Model with Charge Distributions. *Open Science Journal of Modern Physics.* 2014, 1, 6-11.

[37] Malaver, M. Strange Quark Star Model with Quadratic Equation of State. *Frontiers of Mathematics and Its Applications.* 2014, 1, 9-15.

[38] Malaver, M. Charged anisotropic models in a modified Tolman IV space time. *World Scientific News.* 2018, 101, 31-43.

[39] Malaver, M. Charged stellar model with a prescribed form of metric function y(x) in a Tolman VII spacetime. *World Scientific News.* 2018, 108, 41-52.

[40] Malaver, M. Classes of relativistic stars with quadratic equation of state. *World Scientific News.* 2016, 57, 70 -80.

[41] Sunzu, J.; Danford, P. New exact models for anisotropic matter with electric field. *Pramana – J. Phys.* 2017, 89, 44.

[42] Bowers, R.L.; Liang, E.P.T. Anisotropic Spheres in General Relativity. *Astrophys. J.* 1974, 188, 657-665.

[43] Sokolov. A. I. Phase transitions in a superfluid neutron liquid. *Sov. Phys. JETP. 1980,* 52, 575-576.

[44] Bhar P.; Murad MH.; Pant N. Relativistic anisotropic stellar models with Tolman VII spacetime. *Astrophys Space Sci.* 2015, 359: 13. doi: 10.1007/s10509-015-2462-9.

[45] Malaver, M. Some new models of anisotropic compact stars with quadratic equation of state. *World Scientific News.* 2018, 109, 180-194.

[46] Malaver, M. Charged anisotropic matter with modified Tolman IV potential. *Open Science Journal of Modern Physics.* 2015, 2(5), 65-71.

[47] Feroze T.; Siddiqui A. Some Exact Solutions of the Einstein-Maxwell Equations with a Quadratic Equation of State. *J Korean Phys Soc.* 2014, 65, 944-947. doi: 10.3938/jkps.65.944

[48] Sen, B.; Grammenos, Th.; Bhar, P.; Rahaman, F. Mathematical modeling of compact anisotropic relativistic fluid spheres in higher spacetime dimensions. *Math. Meth. Appl. Sci.* 2016, 41, 1062-1067. doi: https://doi.org/10.1002/mma.4128

[49] Kaluza T. Unitätsproblem in der Physik. *Sitz. Preuss. Acad. Wiss.* 1921, Berlin, 966- 972.

[50] Klein O. Quantentheorie und fünfdimensionale Relativitätstheorie. *Z. Physik.* 1926. 37, 895-906.

[51] Paul, B.C.; Dey, S. Relativistic Star in Higher dimensions with Finch and Skea Geometry. *Astrophys Space Sci.* 2018, 363, 220. doi: https://doi.org/10.1007/s10509-018-3438-3.

[52] Bhar, P.; Singh, K.N.; Tello-Ortiz, F. Compact star in Tolman-Kuchowicz spacetime in background of Einstein-Gauss-Bonnet gravity. *Eur. Phys. J. C.* 2019, 79, 922. doi: https://doi.org/10.1140/epjc/s10052-019-7438-4.

[53] Bhar, P.; Govender, M.; Sharma, R. A comparative study between EGB gravity and GTR by modeling compact stars. *Eur. Phys. J. C.* 2017, *77*, 109.

[54] Malaver, M.; Iyer, R.; Kar, A.; Sadhukhan, S.; Upadhyay, S.; Gudekli, E. Buchdahl Spacetime with Compact Body Solution of Charged Fluid and Scalar Field Theory, *arXiv:2204.00981*, General Relativity and Quantum Cosmology, April 2022, 21 pages.

[55] Iyer, R.; Markoulakis, E. Theory of a superluminous vacuum quanta as the fabric of Space, *Phys Astron Int J.* 2021;5(2), 43-53. DOI: 10.15406/paij.2021.05.00233.

[56] Iyer, R. Absolute genesis fire fifth dimension mathematical physics: Absolute genesis physics conjectural mathematics, *Amazon*, 2000, *Engineeringinc.com International Corporation Publisher*, ISBN-13: 978-0970689801, *Kindle Edition*, 2022.

[57] Iyer, R.; O'Neill, C.; Malaver, M.; Hodge, J.; Zhang, W.; Taylor, E. Modeling of Gage Discontinuity Dissipative Physics, *Canadian Journal of Pure and Applied Sciences*, 2022, 16(1), 5367-5377. Publishing Online ISSN: 1920-3853; Print ISSN: 1715-9997.

[58] Buchdahl, H.A. General relativistic fluid spheres. *Phys. Rev.* 116(4), 1959, 1027.

[59] Durgapal, M.C.; Bannerji, R. New analytical stellar model in general relativity. *Phys.Rev. D27*. 1983, 328-331.